\documentclass[traditabstract]{aa}
\usepackage{natbib}
\bibpunct{(}{)}{;}{a}{}{,} % A&A style
\usepackage{txfonts}
\usepackage{graphicx}
\usepackage{url}
\newcommand{\kms}{{\,\rm km\,s}^{-1}}

\def\la{\mathrel{\hbox{\rlap{\hbox{\lower4pt\hbox{$\sim$}}}\hbox{$<$}}}}

\renewcommand{\mag}{\mbox{$\;$mag}}

\begin{document}
% ******************************************************************
\title{New period-luminosity and period-color\\ relations of classical
       Cepheids}
\subtitle{III. Cepheids in SMC}
\author{A. Sandage\inst{1} \and G.~A. Tammann\inst{2} \and B. Reindl\inst{2}}

\institute{%
     The Observatories of the Carnegie Institution of Washington, 
     813 Santa Barbara Street, Pasadena, CA 91101, USA
\and
     Department of Physics and Astronomy, Univ. of Basel
     Klingelbergstrasse 82, 4056 Basel, Switzerland \\
     \email{g-a.tammann@unibas.ch}
}

\date{Received {9 July 2008} / Accepted {\dots}}

% ******************************************************************
%    Abstract
% ******************************************************************
\abstract{
The photometric data for 460 classical, fundamental-mode Cepheids in
the SMC with $\log P > 0.4$ measured by Udalski et~al. have been
analyzed for their period-color (P-C) and period-luminosity (P-L)
relations, and for the variation of amplitude across the instability
strip in a similar way that was done in Papers I and II of this
series. 
The SMC Cepheids are bluer in $(B\!-\!V)^{0}$ at a given period
than for both the Galaxy and the LMC. Their P-C relation in
$(B\!-\!V)^{0}$ is best fit by two lines intersecting at
$P=10\;$days. Their break must necessarily exist also in the P-L
relations in $B$ and/or $V$, but remains hidden in the magnitude
scatter. An additional pronounced break of the P-L relations in $B$,
$V$, and $I$ occurs at $P=2.5\;$days.
The observed slope of the lines of constant period in the HR diagram
agrees with the theoretical expectation from the pulsation equation. 
The largest amplitude Cepheids for periods less than 13 days occur
near the blue edge of the instability strip. The sense is reversed in
the period interval from 13 to 20 days, as in the Galaxy and the LMC.  
The SMC P-L relation is significantly flatter than that for the 
Galaxy, NGC\,3351, NGC\,4321, M31, all of which have nearly the 
same steep slope. The SMC P-L slope is intermediate between that 
of these steep slope cases and the very shallow slope of Cepheids 
in the lower metallicity galaxies of NGC\,3109 and Sextans A/B, 
consistent with the premise that the Cepheid P-L relation varies 
from galaxy-to-galaxy as function of metallicity. 
Failure to take into account the slope differences in the P-L relation
as a function of metallicity using Cepheids as distance indicators
results in incorrect Cepheid distances. Part of the 15\% difference
between our long distance scale -- now independently supported by tip
of the red-giant branch (TRGB) distances -- and that of the HST Key
Project short scale is due to the effect of using an inappropriate P-L
relation.
}
% ******************************************************************
\titlerunning{New P-L and P-C Relations of Classical Cepheids in SMC}
\maketitle
% ******************************************************************

% ******************************************************************
% 1. Introduction
% ******************************************************************
\section{Introduction}
\label{sec:1}
The Galaxy, the LMC, and the SMC are the only three galaxies that 
have reddenings for their Cepheids that are determined by a 
method that is independent of using a fiducial Cepheid 
period-color relation that is incorrectly assumed to be universal. 
Once reliable reddenings are available by such independent methods, 
and by using data for galaxies where the reddening can be assumed 
to be nearly zero, differences in the slope and zero points of P-L 
relations in different galaxies as a function of metallicity 
follow from the observations 
(\citealt*{TSR:08}, hereafter \citeauthor{TSR:08}, see their Fig.~4.; 
\citealt*{ST:08}, hereafter \citeauthor{ST:08}, Fig. 4).  

     In Paper~I of this series (\citealt*{TSR:03}, 
hereafter \citeauthor{TSR:03}) we analyzed the CCD photometric data by 
\citet{Berdnikov:etal:00} for 321 Cepheids in the 
Galaxy for this purpose. A preliminary comparison of the P-L and 
P-C relations for the Cepheids in the LMC (593 stars) and SMC 
(459 stars) was also given there using the CCD photometry by 
\citet{Udalski:etal:99a,Udalski:etal:99b}. 

     In Paper~II (\citealt*{STR:04}, 
hereafter \citeauthor{STR:04}) the lines of constant period that
thread the instability strip were derived for the LMC by correlating
the residuals in magnitude at fixed period read from the P-L relation
with those from the P-C relations for individual Cepheids.
We also derived the difference in the slopes and zero points of the
ridge-lines of the instability strip (i.e. the $\log L$, $\log T_{e}$
HR diagram) for the Galaxy and the LMC, showing that Cepheids in LMC
are hotter at a given luminosity by about 300 K at $\log L = 3.5$ 
(Fig.~20 of \citeauthor{STR:04}), but the size of the difference
varies with period (Fig.~3 of \citeauthor{STR:04}) because of the
slope difference in the two instability strips. 

     The purpose of the present paper is to continue with a 
similar study for the SMC using the data of
\citet{Udalski:etal:99b}. 
The plan of the paper is this. 
The period-color relations in $(B\!-\!V)^{0}$ and $(V\!-\!I)^{0}$ for
SMC in Sect.~\ref{sec:2} are similar to Fig.~6 of \citeauthor{TSR:03}
but with more detail, showing the difference in the P-C relations
between the Galaxy and SMC and emphasizing the break at 10 days for
SMC. 
The slope of the SMC P-L relation in Sect.~\ref{sec:3} is
significantly flatter than those for the Galaxy and LMC, similar to
Fig.~14b of \citeauthor{TSR:03}. 
Lines of constant period in the HR diagram for SMC are derived in
Sect.~\ref{sec:4}. The slopes are compared there with that expected
from the pulsation equation.  
Sect.~\ref{sec:5} shows the correlation of amplitude with position in
the instability strip for various period and absolute magnitude 
ranges. 
The SMC instability strip in $M_{V}$, $(B\!-\!V)^{0}$ and
$(V\!-\!I)^{0}$, with and without the break at 10 days, and the
resulting $\log L_{V}-\log T_{e}$ instability strip is in
Sect.~\ref{sec:6}. 
Comparison of the SMC P-L relation with those in the high metallicity
galaxies of NGC\,3351, NGC\,4321 (with slopes that are nearly
identical to the P-L slope for the Galaxy) and the low metallicity
galaxy NGC\,3109 is in Sect.~\ref{sec:7} showing the significant
difference in the P-L slope as function of metallicity and therefore
that the P-L relation is not universal from galaxy to galaxy.

% ******************************************************************
% 2. Comparison of the P-C and Color-Color Relations for the Galaxy and SMC
% ******************************************************************
\section{Comparison of the P-C and color-color relations for the
         Galaxy and SMC}
\label{sec:2}
%
% ******************************************************************
% 2.1 Observational Data
% ******************************************************************
\subsection{Observational data}
\label{sec:2:1}
In the framework of the OGLE program, \citet{Udalski:etal:99b} have
observed $UBVI$ magnitudes of over 2000 variables in 11 strips
covering much of the central parts of SMC (see their Fig.~1). 
The magnitudes were corrected for absorption using reddening values of
adjacent red-clump stars. Their adopted reddening-to-absorption ratios
are the same for all practical purposes as those used by
\citeauthor{STR:04}. 
The authors derived intensity-averaged mean magnitudes as well as periods
where possible. The light curves in $I$ are exceptionally well defined
by several hundred epochs; the $B$ and $V$ light curves rest on 15 - 40
epochs which is sufficient in almost all cases. The authors have 
derived also light curve amplitudes and Fourier coefficients $R_{21}$
and $\Phi_{21}$. These parameters allowed a convincing separation of
classical Cepheids  from population~II Cepheids and other variables
and to subdivide the classical Cepheids in 1216 fundamental and 833
overtone pulsators. 

     About half of the fundamental-mode Cepheids have periods shorter
than $\log P=0.4$ ($P < 2.5$~days). This great abundance of
short-period Cepheids is specific for SMC with its very low
metallicity of [O/H]$_{\rm T_{e}} = 7.98$ \citep{Sakai:etal:04}. 
Some such Cepheids are known in other metal-poor galaxies, but they
are rare in young populations with higher metallicity as for instance
in LMC. The light curve parameters like magnitude, color, amplitude,
and Fourier coefficients of the short-period Cepheids in SMC are
continuous extensions of those with longer periods. Therefore they are
most likely classical populations~I Cepheids 
\citep[see also][]{Bauer:etal:99,Sharpee:etal:02}, although
taken as a whole they lie below the extrapolated P-L relation defined
by the longer-period Cepheids and they are bluer than the extrapolated
P-C relation defined by the latter. The proliferance of short-period
Cepheids in SMC was explained by \citet{Becker:etal:77} as the effect
of the evolutionary loops of low-mass red giants extending deeper to
the blue side of the CMD, and hence still crossing or at least
penetrating the instability strip, if the metallicity decreases. The
dependence of the loop size on metallicity was first found by
\citet{Hofmeister:67}.  

     In the following we are not concerned with the SMC Cepheids with
$\log P<0.4$ because our emphasis lies on the difference of Cepheids
from galaxy to galaxy and on the ensuing effect on distance
determinations. For this purpose the short-period Cepheids are not
helpful because statistically meaningful samples of Cepheids with 
$\log P<0.4$ are not available in other galaxies. Therefore only SMC
Cepheids with $\log P>0.4$ are considered in the following.  

     The sample of fundamental-mode SMC Cepheids with $\log P>0.4$ was
cleaned by \citet{Udalski:etal:99b} by removing all objects deviating
by $2.5\sigma$ or more from a common P-L relation. This left 469
Cepheids with $V$ magnitudes. We have removed some additional outliers
resulting in 460 Cepheids with $V$ and $I$ magnitudes, of which 439
objects have also $B$ magnitudes. Their data are the basis of the
following analysis.

% ******************************************************************
% 2.2 The Period-Color Relation
% ******************************************************************
\subsection{The period-color Relation}
\label{sec:2:2}
We adopt the individual reddenings for the SMC Cepheids derived 
by \citet{Udalski:etal:99b} determined from the observed color of the
red clump stars in the vicinities of the Cepheids. It is important to
note that these reddenings have been determined by a method that is
independent of using some adopted fiducial P-C relation. 
Because the P-C color relations vary with metallicity at the $0.2\mag$
level, they will vary in slope and zero point from galaxy-to-galaxy
for different metallicities. Systematic errors in the calculated
reddenings will be made if a fixed ``universal'' fiducial P-C relation
is used in galaxies with metallicities that differ from that of the
fiducial sample.  

     Least-squares solutions of the $(B\!-\!V)^{0}$ and
$(V\!-\!I)^{0}$ colors of all Cepheids with $\log P>0.4$ selected from
the SMC sample of \citeauthor{Udalski:etal:99b} yield
\begin{eqnarray}
\label{eq:PC:BV}
  (B\!-\!V)^0 &=& (0.360\!\pm\!0.016)\log P + (0.235\!\pm\!0.012),\; \sigma=0.09,
\\
\label{eq:PC:VI}
  (V\!-\!I)^0 &=& (0.276\!\pm\!0.014)\log P + (0.450\!\pm\!0.010),\; \sigma=0.08.
\end{eqnarray}
(Flatter slopes for SMC are shown in Table~4 of \citeauthor{TSR:08},
but they do actually include also the many Cepheids with $\log P<0.4$).

     For comparison the Galactic P-C relations for $\log P>0.4$ are
repeated here from \citeauthor{TSR:03} (the update here in the
equations for the Galaxy compared with the Galactic equations in
\citeauthor{TSR:03} and \citeauthor{STR:04}, obtained by subtracting
Eqs.~(16)--(18) there, is insignificant for this comparison):
\begin{eqnarray}
\label{eq:PC:BV:gal}
  (B\!-\!V)^0 &=& (0.366\!\pm\!0.015)\log P + (0.361\!\pm\!0.013),\; \sigma=0.07,
\\
\label{eq:PC:VI:gal}
  (V\!-\!I)^0 &=& (0.256\!\pm\!0.017)\log P + (0.497\!\pm\!0.016),\; \sigma=0.07.
\end{eqnarray}
The P-C relations of SMC and the Galaxy have quite similar slopes
above $\log P=0.4$, but the striking difference is that the SMC
Cepheids are bluer by $\Delta(B\!-\!V)^{0}=0.13$ and
$\Delta(V\!-\!I)^{0}=0.03$ than the Galactic ones. This was discovered
already by \citet[][see also
\citealt{Gascoigne:74}]{Gascoigne:Kron:65} and made definitive by  
\citet{Laney:Stobie:86,Laney:Stobie:94} as a temperature difference.

\begin{figure}
   \resizebox{\hsize}{!}{\includegraphics{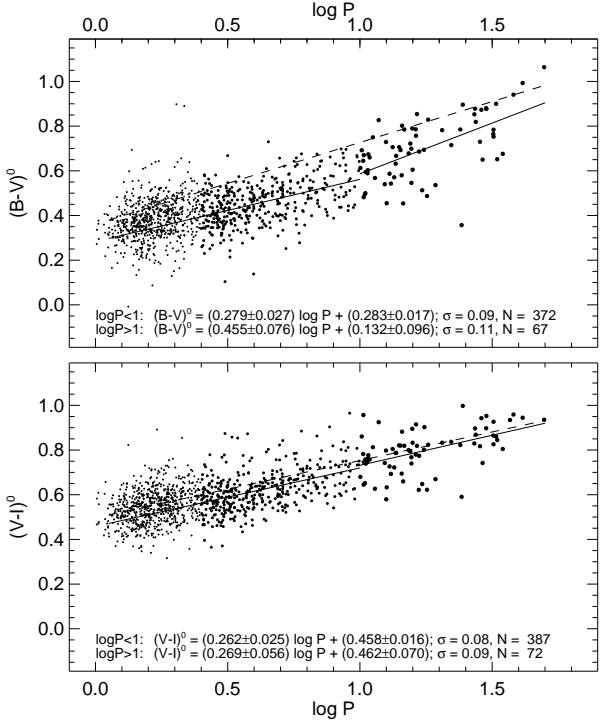}}
    \caption{The SMC period-color relations in $(B\!-\!V)^{0}$ and
   $(V\!-\!I)^{0}$. Small dots are for $\log P<0.4$. Large dots are
   for $\log P>1.0$, separated at this period to emphasize the break
   at that period. The least squares equations for $\log P>0.4$ and
   assuming a break at 10 days (solid lines) are listed
   inside the diagram. The P-C relations for the Galaxy (no-break)
   are the dashed lines taken from Eqs.~(\ref{eq:PC:BV:gal}) and
   (\ref{eq:PC:VI:gal}). The period-color relations for the LMC (not
   shown) are intermediate between the Galaxy and the SMC as given by
   Eqs.~(2) and (3) for $(B\!-\!V)^{0}$ and Eqs.~(5) and (6) for
   $(V\!-\!I)^{0}$ in \citeauthor{STR:04}.} 
    \label{fig:01}
\end{figure}

     The individual Cepheids defining the SMC P-C relations of 
Eqs.~(\ref{eq:PC:BV}) and (\ref{eq:PC:VI}) are plotted in
Fig.~\ref{fig:01}. Here the $(B\!-\!V)^{0}$ data strongly suggest that
a single-slope P-C relation does not provide an optimum fit, but that
the slope changes around $\log P\sim1$. Two separate fits for the
period intervals $\log \gtrless1$ lead for $0.4<\log P<1$ to
\begin{equation}
\label{eq:PC:BV:lt1}
  (B\!-\!V)^0 = (0.279\!\pm\!0.027)\log P + (0.283\!\pm\!0.017),\; \sigma=0.09,
\end{equation}
and for $\log P>1$ to
\begin{equation}
\label{eq:PC:BV:ge1}
  (B\!-\!V)^0 = (0.455\!\pm\!0.076)\log P + (0.132\!\pm\!0.096),\; \sigma=0.11.
\end{equation}
The change of slope has here a significance of $2\sigma$. The case is
reminiscent of LMC where the break at $P=10^{\rm d}$ is well
documented. In both galaxies the long-period Cepheids have a
steeper color gradient than at shorter periods.

% ********************************************************
% Figure 1: SMC - PC
% ********************************************************
% ********************************************************

     For the sake of completeness we also give the separate SMC P-C
relations in $(V\!-\!I)^{0}$ for the period interval $0.4< \log P<1$: 
\begin{equation}
\label{eq:PC:VI:lt1}
  (V\!-\!I)^0 = (0.262\!\pm\!0.025)\log P + (0.458\!\pm\!0.016),\; \sigma=0.08,
\end{equation}
and for $\log P>1$: 
\begin{equation}
\label{eq:PC:VI:ge1}
  (V\!-\!I)^0 = (0.269\!\pm\!0.056)\log P + (0.462\!\pm\!0.070),\; \sigma=0.09.
\end{equation}
They turn out to be statistically undistinguishable. Their slope is
also quite close to the Galactic slope in Eq.~(\ref{eq:PC:VI:gal}).

     A comparison of the Galactic P-C relation in $(B\!-\!V)^{0}$ in 
Eq.~(\ref{eq:PC:BV:gal}) with the corresponding SMC P-C relations in
Eqs.~(\ref{eq:PC:BV:lt1}) and (\ref{eq:PC:BV:ge1}) reveals a slope
difference, both above and below the 10 day break. It will be shown in
Sect.~\ref{sec:5} that these slope differences translate into slope
differences in temperature in the $M_{V}-\log T_{e}$ instability
strip diagram (Figs.~\ref{fig:08} and \ref{fig:09} later).

% ******************************************************************
% 2.3 Proof via the (B-V)o,(V-I)o Color-Color Diagram that the 
%     intrinsic color difference between SMC and [via?] the Galaxy is real 
% ******************************************************************
\subsection{Proof via the $(B\!-\!V)^{0}$, $(V\!-\!I)^{0}$ color-color
  diagram that the intrinsic color difference between SMC and the
  Galaxy is real} 
\label{sec:2:3}
The validity of the color differences between the Galaxy and 
SMC in Fig.~\ref{fig:01} depends, of course, on the accuracy of the 
reddening determinations for both the Galaxy and the SMC. One 
might suppose that the observed color difference in Fig.~\ref{fig:01}
is not real because of errors in the reddenings, although the 
methods used by \citet{Fernie:90,Fernie:94} and \citet{Fernie:etal:95}
are very powerful. However, this possibility can be disproved by 
appeal to the $(B\!-\!V)^{0}$, $(V\!-\!I)^{0}$ color-color diagram for
SMC compared with that for the Galaxy shown in Fig.~7a of
\citeauthor{TSR:03}. There is an offset in that diagram of
$\sim\!0.1\mag$ in $(B\!-\!V)^{0}$ at a given $(V\!-\!I)^{0}$,  
with SMC Cepheids being bluer than those in the Galaxy. Because 
the reddening line is nearly parallel to the observed correlation 
line (Panel~a of Fig.~7a of \citeauthor{TSR:03}), errors in the
reddenings will not produce the observed color offsets seen in
Panel~c. Furthermore, the observed offset of SMC relative to the
Galaxy is close to what is expected from model atmosphere calculations
by Bell \& Tripicco from Table~6 of \citet*[][hereafter
\citeauthor{SBT:99}]{SBT:99}, shown in Panel~d of Fig.~7a in
\citeauthor{TSR:03}. Hence, the color offset in Fig.~\ref{fig:01} here
of SMC relative to the Galaxy is not due to reddening errors, but is
real.

% ******************************************************************
% 3. The SMC P-L Relation Compared With the Galaxy and the LMC
% ******************************************************************
\section{The SMC P-L relation compared with the Galaxy and the LMC}
\label{sec:3}
\begin{figure}
   \vskip0.5cm
   \resizebox{\hsize}{!}{\includegraphics{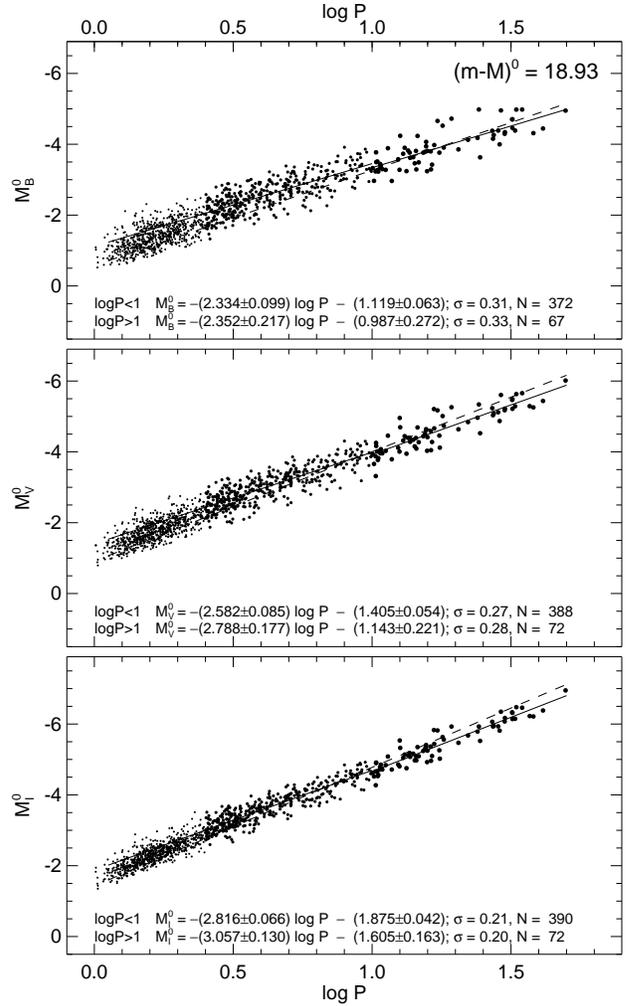}}
    \caption{The P-L relations in $B$, $V$, and $I$ for the SMC from
   the absorption corrected data of \citet{Udalski:etal:99b}. The
   different symbols for $\log P<0.4$ and $\log P>1.0$ are the same as
   in Fig.~\ref{fig:01}. The least-squares solutions for
   $0.4<\log P<1.0$ and $\log P>1.0$, assuming a break at $10^{\rm d}$
   (solid lines) are listed inside the diagram. The P-L relations for
   the Galaxy from Eqs.~(16)--(18) of \citeauthor{STR:04} are shown
   for comparison as the dashed line in each panel.} 
    \label{fig:02}
\end{figure}

The SMC P-L relations in $B$, $V$, and $I$ from the data by 
\citet{Udalski:etal:99b} are shown in Fig.~\ref{fig:02}, using a zero
point for the SMC distance modulus of $(m-M)^{0} =18.93$ taken from a
summary given elsewhere for modulus values determined after 2004 
(\citeauthor{TSR:08}, Table~7). The symbols are the same as in
Fig.~\ref{fig:01}. 

% ********************************************************
% Figure 2: SMC - PL
% ********************************************************
% ********************************************************

     Least squares solutions for the ridge-line P-L equations were
made with and without a break at 10 days using Cepheids with 
$\log P > 0.4$. This period restriction is to make the comparison here
with LMC more secure where the same period cut-off was used. The break
equations are shown within the borders of Fig.~\ref{fig:02}. 
The non-break equations are 
\begin{eqnarray}
 \label{eq:PL:B}
   M^{0}_{B} & = & -(2.222\!\pm\!0.054)\log P - (1.182\!\pm\!0.041),\;
   \sigma=0.31 \\
 \label{eq:PL:V}
   M^{0}_{V} & = & -(2.588\!\pm\!0.045)\log P - (1.400\!\pm\!0.035),\;
   \sigma=0.27 \\
 \label{eq:PL:I}
   M^{0}_{I} & = & -(2.862\!\pm\!0.035)\log P - (1.847\!\pm\!0.027),\;
   \sigma=0.21.
\end{eqnarray}
The break is less significant for SMC Cepheids than in the LMC.

     The P-L relations for the Galaxy from Eqs. (16)--(18) of 
\citeauthor{STR:04} are shown in Fig.~\ref{fig:02} by the dashed
line. The deviation of the SMC P-L relation from the Galaxy is
clear. The SMC Cepheids at $\log P= 0.4$ are $\sim\!0.2\mag$ brighter
in $V$ than those in the Galaxy at $\log P= 0.4$ and are
$\sim\!0.3\mag$ fainter at $\log P=1.5$.  
The difference is larger in $B$ and smaller in $I$.       

     If one distrusts the Galaxy P-L slopes and zero points, 
either because of questions concerning main sequence fittings and 
problems with absorption \citep{vanLeeuwen:etal:07} for star 
clusters and associations, or because of doubts about the 
reliability of the Baade/Becker/Wesselink moving atmosphere 
results due to an uncertainty on the velocity projection factor 
\citep{Fouque:etal:07}, comparison of the over-all slopes of the SMC
P-L relations with those of the LMC again shows a difference, with 
the SMC slope being flatter in all three colors. The significance of
the slope difference decreases from $B$ to $I$. To avoid confusion,
the LMC P-L relation is not overlayed in Fig.~\ref{fig:02} to show
this, but the LMC P-L Eqs.~(7)--(9) of \citeauthor{STR:04} differ
significantly from the equations set out in Fig.~\ref{fig:02} for SMC. 
The over-all SMC P-L relation compared with LMC is shown
later in Fig.~\ref{fig:10} in Sect.~\ref{sec:7}.

The comparison of the P-L relations of SMC and LMC becomes more
complicated if one allows for the break at $P=10^{\rm d}$. In the
interval $0.4<\log P<1.0$ the SMC slopes are considerably 
{\em flatter\/} than those of LMC. The difference is again more
pronounced in $B$ than in $I$ and intermediate in $V$. On the other
hand the SMC slopes are {\em steeper\/} for $\log P>1.0$ than those in
LMC, the difference being about the same in all three colors.

     A very tight P-L relation of SMC is obtained when the individual
Cepheids are slid along the constant-period lines onto the ridge line
of the P-L relation. The procedure is explained in \citeauthor{TSR:08}
(Eq.~(3)) and the result is illustrated in Fig.~3b there. Because of
the small dispersion of $0.15\mag$ the P-L relation is very well
defined. The break at $\log P=1$ is less pronounced than in LMC, 
but still has a significance of $2.3\sigma$ and is of opposite sign!
Any claim that the P-L relations of LMC and SMC in Fig.~3a and 3b of
\citeauthor{TSR:08} were the same would be a denial of the evidence.

% ******************************************************************
% 4. Lines of Constant Period in the SMC Instability Strip 
% ******************************************************************
\section{Lines of constant period in the SMC instability strip}
\label{sec:4}
%
% ******************************************************************
% 4.1 Empirical Determination From the Data Directly
% ******************************************************************
\subsection{Empirical determination from the data directly}
\label{sec:4:1}
Lines of constant period in the HR diagram thread from the luminous,
high temperature upper left to the low luminosity, low temperature
lower right. Therefore, Cepheids of a given period near the blue edge 
of the instability strip will be brighter than Cepheids near the red
edge. Hence, there should be a correlation between luminosity
residuals in the P-L diagram and the color residuals in the P-C
diagram at fixed periods. 
The slope $\beta$ of the constant-period lines in the instability
strip in the $M, (B\!-\!V)^{0}$ plane is given by 
$\beta=\Delta M_{BVI}/\Delta(B-V)^{0}$.

\begin{figure*}
   \sidecaption
   \resizebox{0.63\hsize}{!}{\includegraphics{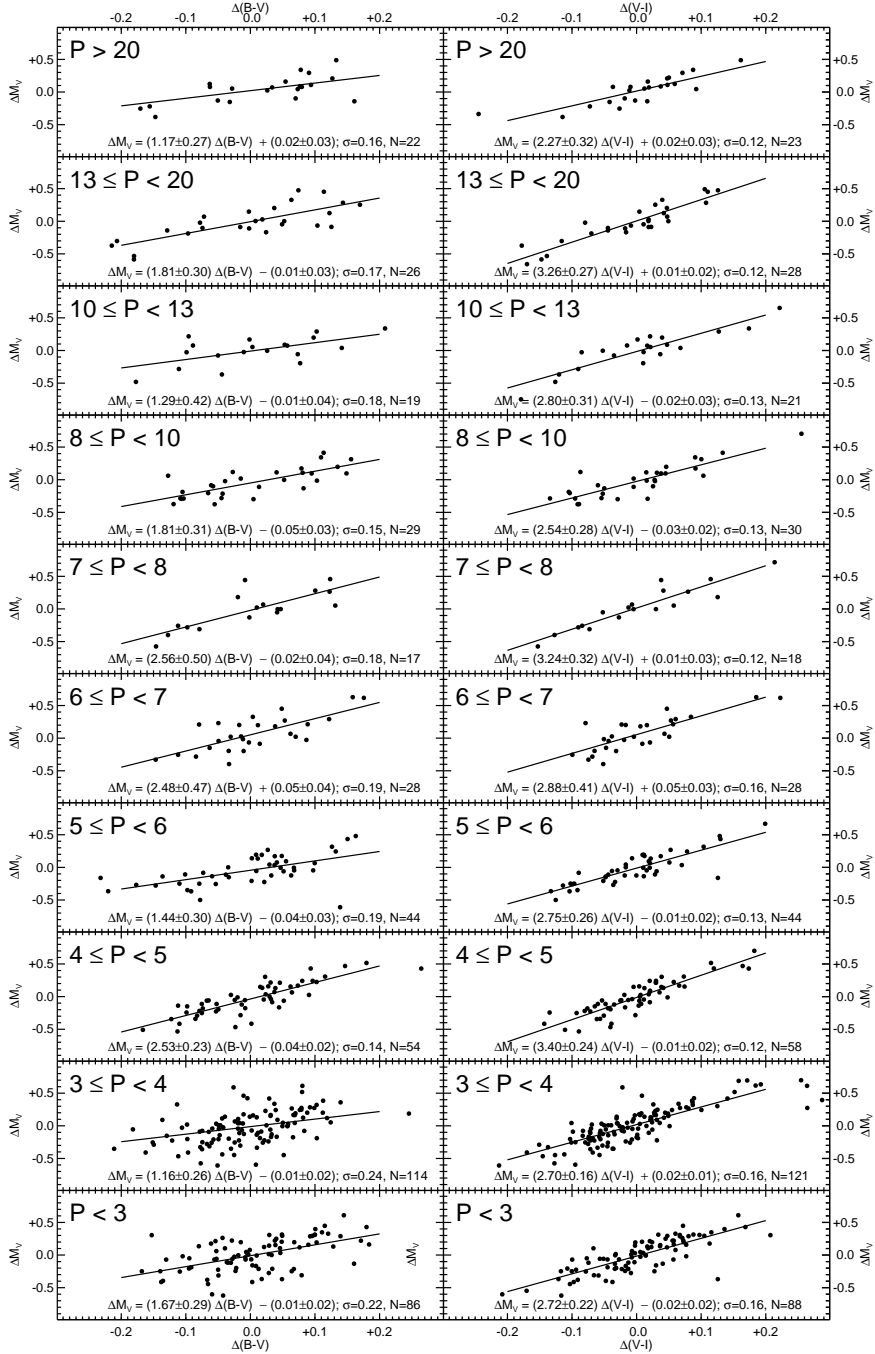}}
    \caption{The slope of the lines of constant period using the
   ratios of the residuals in $(B\!-\!V)^{0}$ and $(V\!-\!I)^{0}$ from
   the P-C relations of Fig.~\ref{fig:01}, and the magnitude residuals
   from Fig.~\ref{fig:02} for the Cepheids in SMC in the period
   intervals indicated. The sense of the correlations is that Cepheids
   near the blue border of the instability strip are intrinsically
   brighter than those near the red border.} %\vskip0.8cm}
    \label{fig:03}
\end{figure*}

     The constant-period lines were well determined for the LMC, with
average slopes of 
$\langle\beta_{B-V}\rangle=1.8$ and 
$\langle\beta_{V-I}\rangle=2.5$ and no significant variation with
period (\citeauthor{STR:04}, Fig.~9 and Table~4). 
The same determination was made there for the SMC and is repeated here
with the results shown in Fig.~\ref{fig:03}, binned in small intervals
of period. The slopes at given period intervals are similar to those
for LMC. The mean slopes over all periods are
$\langle\beta_{B-V}\rangle=1.8\pm0.2$ and  
$\langle\beta_{V-I}\rangle=2.8\pm0.1$ for SMC. 

% ********************************************************
% Figure 3: SMC - CPL *** two-column-figure
% ********************************************************
% ********************************************************

% ******************************************************************
% 4.2 Predicted Slope of the Constant Period Lines Using the Pulsation Equation
% ******************************************************************
\subsection{Predicted slope of the constant period lines using the 
  pulsation equation}
\label{sec:4:2}
The equation of the family of constant period lines in the 
$L_{\rm bol}$, $T_{e}$ HR diagram follows from the 
$P(L,\mbox{mass},T_{e})$ pulsation equation when the mass is
eliminated using a mass-luminosity relation for the Cepheids. It is
shown elsewhere (\citeauthor{ST:08}) that the resulting equation for
the family is    
\begin{equation}
     \log L_{\rm bol}=5.472 \log T_{e} + 1.572 \log P - 18.406. 
\label{eq:01}
\end{equation}
Changing into a $M_{V}, (B\!-\!V)$ dependence by neglecting the
bolometric correction (using the approximation that 
$M_{\rm bol}=M_{V}$, valid to $\sim0.1\mag$ over much of the color
range of the Cepheids), and using a linearized $(V\!-\!I)$,
temperature relation by interpolating in Table~6 of
\citeauthor{SBT:99} for the $\log g$ of Cepheids and with 
[A/H]$=-0.7$ for the SMC Cepheids gives 
\begin{equation}
       \log T_{e} =  -0.24 (V\!-\!I) + 3.924,              
\label{eq:02}
\end{equation}
over the range $0.5<(V\!-\!I)<0.9$. Substituting into
Eq.~(\ref{eq:01}) gives the equation of the constant period lines as, 
\begin{equation}
     M_{V} = 3.32 (V\!-\!I) - 3.93 \log P -2.92,            
\label{eq:03}
\end{equation}
where we have used $M_{V} = -2.5 \log L + 4.75$. 

     The predicted slope here of 3.3 for the constant period 
lines in $V\!-\!I$ agrees well enough with the observed slope of 
$2.8\pm0.1$ considering the approximations we have made. 
A similar agreement obtains in $B\!-\!V$. 
These agreements to within the approximations show that the observed
lines of constant period in Fig.~\ref{fig:03} are understood as a
consequence of the Ritter $P\sqrt{\rho}$ pulsation condition.

% ******************************************************************
% 5. Correlation of Amplitude With position in the Instability Strip 
% ******************************************************************
\section{Correlation of amplitude with position in the instability strip}
\label{sec:5}
By analogy with the more easily visualized case of the RR~Lyrae stars
in globular clusters where the horizontal branch cuts the instability
strip at nearly constant $M_{V}$, we make the same correlations for
the Cepheids here as is the custom for globular cluster variables. The
strong correlation of amplitude with strip position in globular
clusters at fixed absolute magnitude (called now the Bailey diagram)
shows that the largest amplitudes for RR~Lyrae stars occur near the
blue edge, decreasing monotonically toward the red edge.  

     However, the situation is more complicated for the Cepheids 
because there is no restriction in absolute magnitude as with the 
RR~Lyrae stars. The absolute magnitude in $V$ does not change
appreciably for globular cluster variables, hence there is no
period-luminosity relation in $V$. Of course, in other bands than $V$
where the horizontal branch is not horizontal, there is a ``pseudo''
P-L relation caused however by the change of luminosity with color
across the strip whereas for Cepheids the vertical change of absolute
magnitude with period in the P-L strip is the principal effect.

     Said differently, for RR~Lyrae stars the principal parameters
describing the fine structure of the instability strip are period,
color, and amplitude, not luminosity, whereas for Cepheids the
principal parameters are period and absolute magnitude, with color and
amplitude having only a secondary role. It is the cutting of the strip
by the horizontal branch in the globular clusters that restricts the
way the parameters appear in the data.  

     The color-amplitude correlation for the RR~Lyrae data is the
striking feature of the data for the cluster variables. This
translates into the more easily observed period-amplitude Bailey
diagram which often dominates the RR~Lyrae analyses. The correlation
is less noticeable for the Cepheids because of the dominance of the
overwhelming luminosity variation with period that is absent in the
cluster variables. 

      Nevertheless there is a color-amplitude effect at fixed 
period (or fixed absolute magnitude) for the Cepheids, with 
the highest amplitude Cepheids occurring at the blue edge of the 
strip, just as for the RR~Lyrae stars, at least in the period 
range from 3 to 10 days in the LMC (\citeauthor{STR:04}, Fig.~11) with
the trend reversing for periods from 10 to 20 days, and perhaps 
returning again to high amplitudes at the blue edge for 
$P>20\;$days (\citeauthor{STR:04}, Fig.~11). 
The effect was seen initially in the Galaxy where the reversal of the
sense occurs at $\log P=0.86$, and returns to the original sense for
$\log P>1.3$. This result was originally suggested from a small sample
of Galaxy Cepheids \citep{ST:71}, and was confirmed in
\citeauthor{STR:04} (Sect.~6.3) from the much larger sample of Galaxy
Cepheids in the \citet{Berdnikov:etal:00} sample. 

     The effect was shown to be the same for the LMC Cepheids in 
the period range from 3 to 10 days and for $P>20\;$days, and again 
the sense is reversed for $10<P<20\;$days 
(\citeauthor{STR:04}, Figs.~11 and 12).  

\begin{figure}
   \centering
   \resizebox{0.95\hsize}{!}{\includegraphics{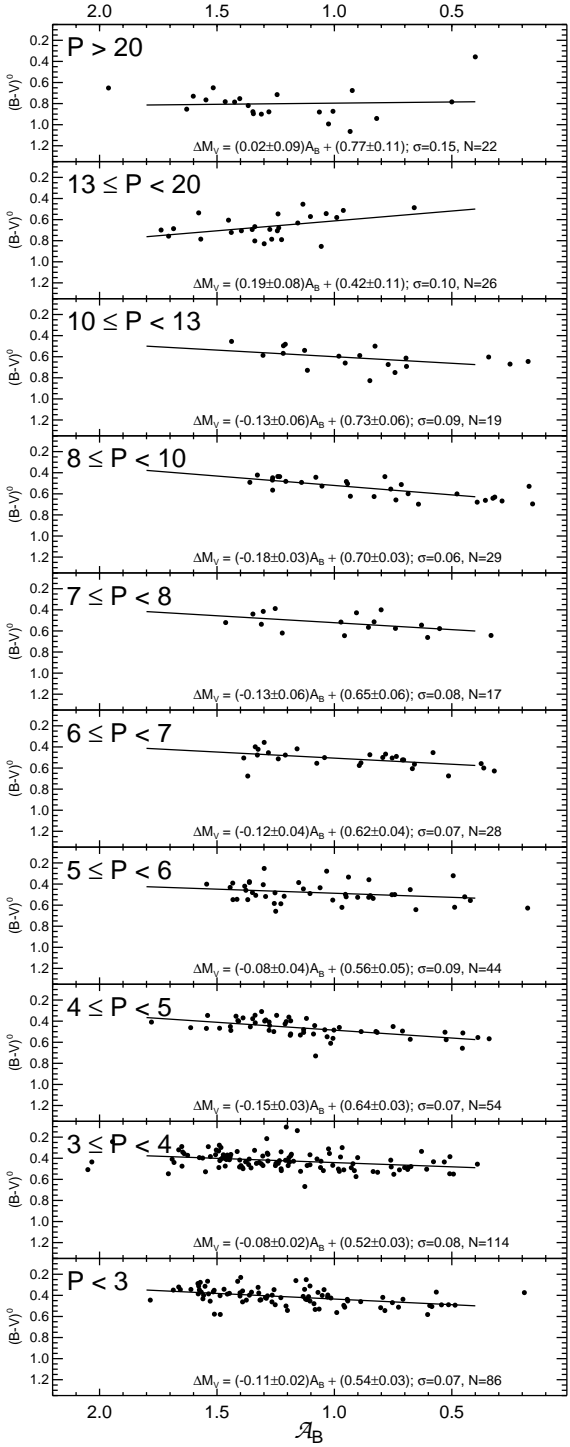}}
    \caption{Correlation of $B$ band amplitude with color for SMC
   Cepheids in various period ranges, showing that the largest
   amplitude variables are near the blue edge of the instability strip 
   except in the period range $13<P<20$~days.} 
    \label{fig:04}
\end{figure}

\begin{figure}
   \centering
   \resizebox{0.95\hsize}{!}{\includegraphics{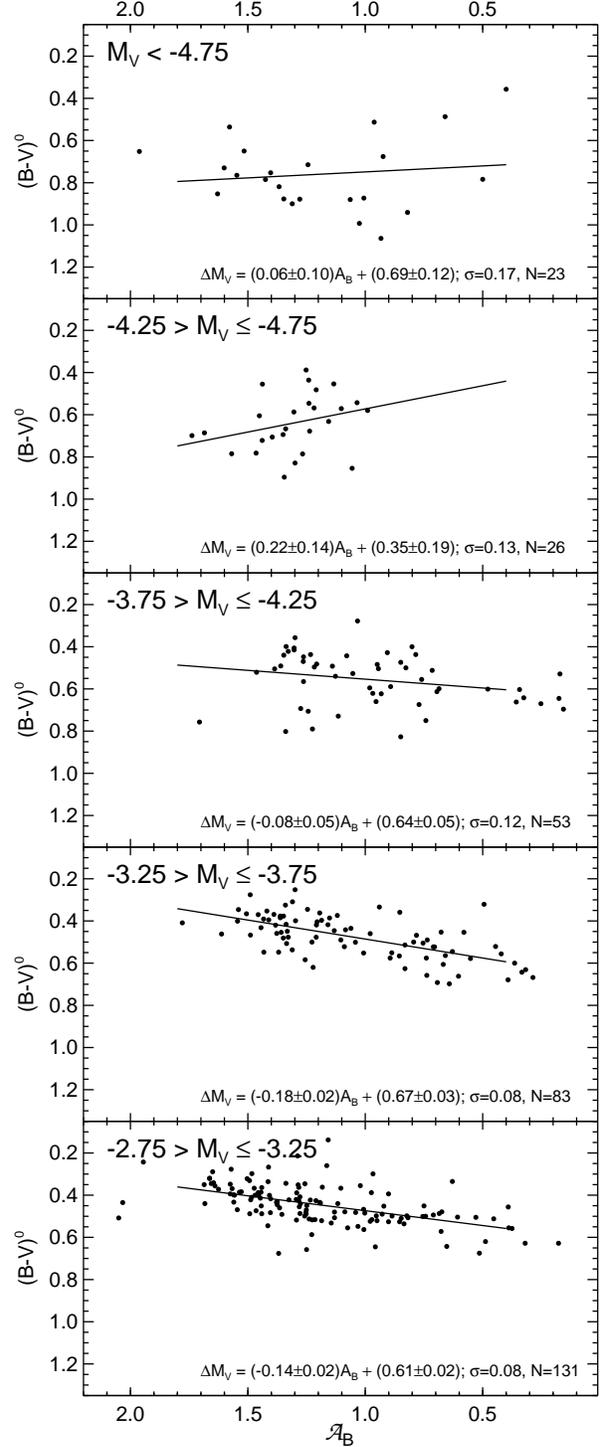}}
    \caption{Same as Fig.\ref{fig:04} but binned in absolute magnitude
   intervals rather than period intervals. The reversal in the sense
   of the correlation of amplitude with color for the absolute
   magnitude interval near $M_{V}\sim-4.5$ is the same as in the LMC.} 
    \label{fig:05}
\end{figure}

     We have analyzed the large data sample for the SMC in the 
same way, with the results shown in Figs.~\ref{fig:04} and
\ref{fig:05}. The sense of the trends are identical with those in the
Galaxy and LMC. The reversal of the sense for periods between 10 and
20 days is also evident here for the SMC although the period range is
different. For SMC the reversal occurs between 13 and 20 days, while
in the Galaxy it is between 7 and 20 days and for the LMC it is
between 10 and 20 days. The absolute magnitude range for the change of
sense is almost identical in each case at between $M_{V}$ of $-4.25$
and $-4.75$.
The reason for this behavior is not yet understood.

% ********************************************************
% Figure 4: SMC - Bamp.P
% ********************************************************
% ********************************************************
% ********************************************************
% Figure 5: SMC - Bamp.Mv
% ********************************************************
% ********************************************************

     Because the amplitudes are largest at the blue edge of the
instability strip (except for the period range of 10 to 20 days),  
and because the blue edge of the strip is brighter than the red edge
at fixed period (Fig.~\ref{fig:03} here and Fig.~9 of
\citeauthor{STR:04} for LMC), there must be a period, magnitude,
amplitude correlation in the sense that large amplitude Cepheids are
brightest, seen in Eqs.~(36)--(41) of \citeauthor{STR:04}. 
The maximum effect is at the level of $\sim\!0.2\mag$, and is a
function of the ``amplitude defect'', a parameter introduced by  
\citet{Kraft:60}, calculated for each Cepheid from the deviation of 
a given amplitude from the upper envelope line of the 
period-amplitude relation. The bias effect on distances and the
severity of the effect for the extragalactic distance scale is small
as discussed in detail in \citeauthor{STR:04}, Sect.~6.3, and is not
repeated here. Nevertheless, the existence of the effect leads to the
notion of fine structure in the P-L relation depending on amplitude,
and is to be discussed separately elsewhere.            

% ********************************************************
% Figure 6: SMC - CMD with break *** two-column-figure
% ********************************************************
\begin{figure}
   \centering
   \resizebox{1.05\hsize}{!}{\includegraphics{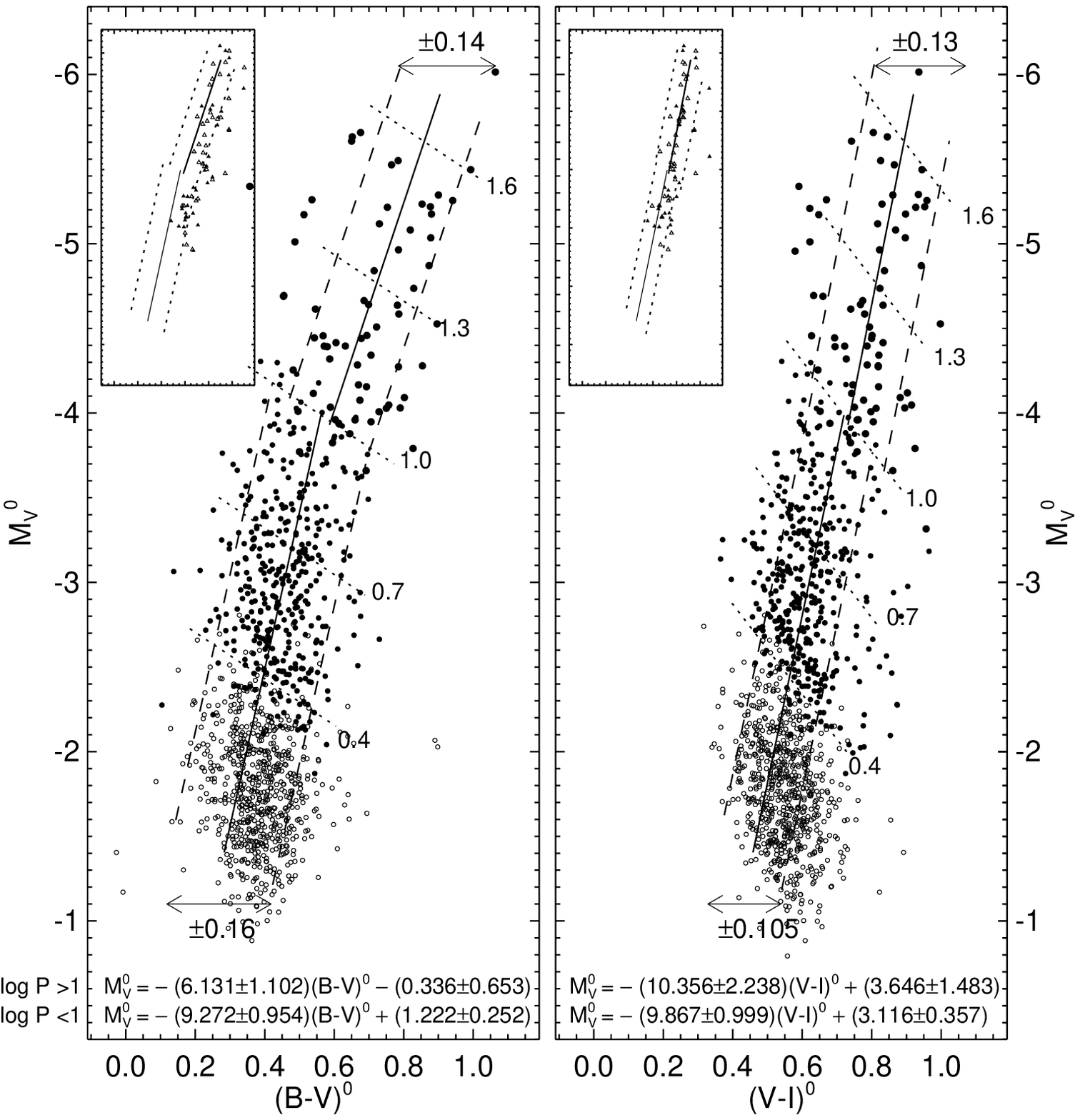}}
    \caption{The Cepheid instability strip for the SMC Cepheids in 
   the $M_{V}$-color plane allowing for a break at 10 days (see
   text). Variables with periods less than $\log P=0.4$ are shown as
   small open circles. Five lines of constant period are drawn and
   marked by their log P values. The insert diagrams repeat the ridge
   line equations for SMC. The 53 individual Cepheids in the Galaxy
   from the data in \citeauthor{TSR:03} and \citeauthor{STR:04} are
   shown as points in the inserts. The significant blueward offset of
   the SMC Cepheids in the $(B\!-\!V)^{0}$ insert panel is evident.} 
    \label{fig:06}
\end{figure}
% ********************************************************

% ******************************************************************
% 6. SMC Instability Strip in Color and in Te  
% ******************************************************************
\section{SMC instability strip in color and in \boldmath $T_{e}$}
\label{sec:6}
%
% ******************************************************************
% 6.1 The Magnitude-Color Instability Strip
% ******************************************************************
\subsection{The magnitude-color instability strip}
\label{sec:6:1}
The instability strip in $M_{V}$ vs. $(B\!-\!V)^{0}$ and
$(V\!-\!I)^{0}$ for the SMC Cepheids is in Fig.~\ref{fig:06}. 
The ridge line equations with a break at 10 days, shown at the bottom
of Fig.~\ref{fig:06}, are obtained by substituting the $\log P$ term
in the P-L relations at the bottom of Fig.~\ref{fig:02} by means of
the P-C relations given in Eqs.~(\ref{eq:PC:BV:lt1})--(\ref{eq:PC:VI:ge1}).
The ridge lines are shown as full lines in Fig.~\ref{fig:06}.
The break is suggestive in the $(B\!-\!V)^{0}$ panel but is not as
pronounced in $(V\!-\!I)^{0}$. The blue and red edges of the strip,
shown as dashed lines, are put parallel to the ridge lines using 
widths of 0.16 and $0.105\mag$, respectively, for $P<10^{\rm d}$ and
0.14 and $0.13\mag$ for $P>10^{\rm d}$. These values are derived from
the calculated rms spreads in the P-L relations in Fig.~\ref{fig:02}
and the mean slopes of the lines of constant period from
Sect.~\ref{sec:4}. The lines of constant period are shown for 
$\log P$ of 0.4, 0.7, 1.0, 1.3 and 1.6 using the formulation 
in Sect.~\ref{sec:4}.

     The insert diagrams in Fig.~\ref{fig:06} repeat the ridge-line
equations and are shown as the solid lines with the edges of the
instability strip also repeated. The individual points in the inserts
are the 53 Cepheids in the Galaxy sample taken from Tables~3 and 4 of 
\citeauthor{TSR:03}, as updated in \citeauthor{STR:04}, Sect.~4.2. The
difference in the SMC ridge lines and the position of the Cepheids in
the Galaxy, is evident. This difference is the same as seen in
Fig.~\ref{fig:01} but in a different representation.       

\begin{figure}
   \centering
   \resizebox{1.05\hsize}{!}{\includegraphics{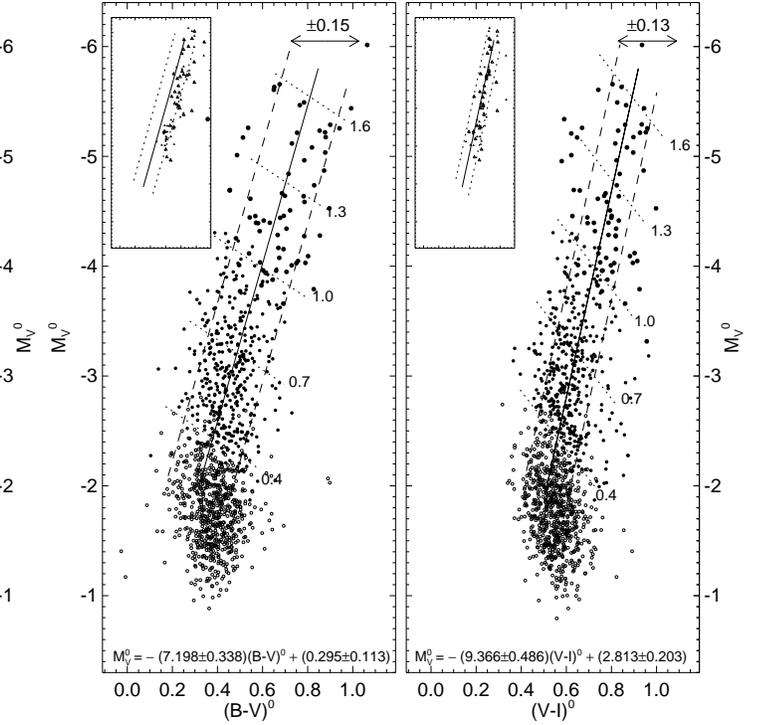}}
     \caption{Same as Fig.~\ref{fig:06} but the ridge-line solutions
   are made with no break at 10 days. The insert diagrams again
   emphasize the difference in the instability strips in SMC (solid
   lines) compared with the Galaxy (individual points) in both slope
   and zero point.}
    \label{fig:07}
\end{figure}

     Fig.~\ref{fig:07} is the same as Fig.~\ref{fig:06} but with 
no allowance for the break at 10 days, i.e.\ the P-L relations in 
Eqs.~(\ref{eq:PL:B})--(\ref{eq:PL:I}) were combined with the P-C
relations in Eqs.~(\ref{eq:PC:BV}) and (\ref{eq:PC:VI}) to yield the
no-break ridge line equations at the bottom of Fig.~\ref{fig:07}.
Again, the differences in the slope and the zero point of the 
instability strip ridge lines for SMC compared with the Galaxy 
Cepheids (the individual points) are evident in the insert diagrams.

% ********************************************************
% Figure 7: SMC - CMD no break *** two-column-figure?
% ********************************************************
% ********************************************************

% ******************************************************************
% 6.2 The Instability Strip in log LV vs. log Te
% ******************************************************************
\subsection{The instability strip in $\log L_{V}$ vs. $\log T_{e}$}
\label{sec:6:2}
The data in Figs.~\ref{fig:06} and \ref{fig:07} have been transformed
to the $\log L_{V}$, $\log T_{e}$ representation of the instability
strip in the same way as we did for the Cepheids in LMC in Fig.~20 of
\citeauthor{STR:04}. The $(B\!-\!V)^{0}$ and $(V\!-\!I)^{0}$ colors
are changed into $\log T_{e}$ by interpolations in Table~6 of
\citeauthor{SBT:99} using appropriate $\log g$ values as function of
period\footnote{%
$\log g=-1.09\log P+2.64$ as derived in \citeauthor{STR:04} (Eq.~49).} 
(from the P-L relation and the assumption of the Cepheid
mass-luminosity relation), [A/H] metallicity values ($0.0$ for the
Galaxy, $-0.5$ for LMC, and $-0.7$ for SMC), and a turbulent velocity
of $1.7\kms$. The two values of $\log T_{e}$ for each Cepheid from the
colors were averaged. We also assumed that the bolometric correction
was small enough to be neglected in converting $M_{V}$ to $\log L$ by 
$\log L_{V}=-0.4M_{V}+1.9$ (using $M_{\rm bol}(\mbox{sun}) = 4.75$).  

\begin{figure}
   \centering
   \resizebox{\hsize}{!}{\includegraphics{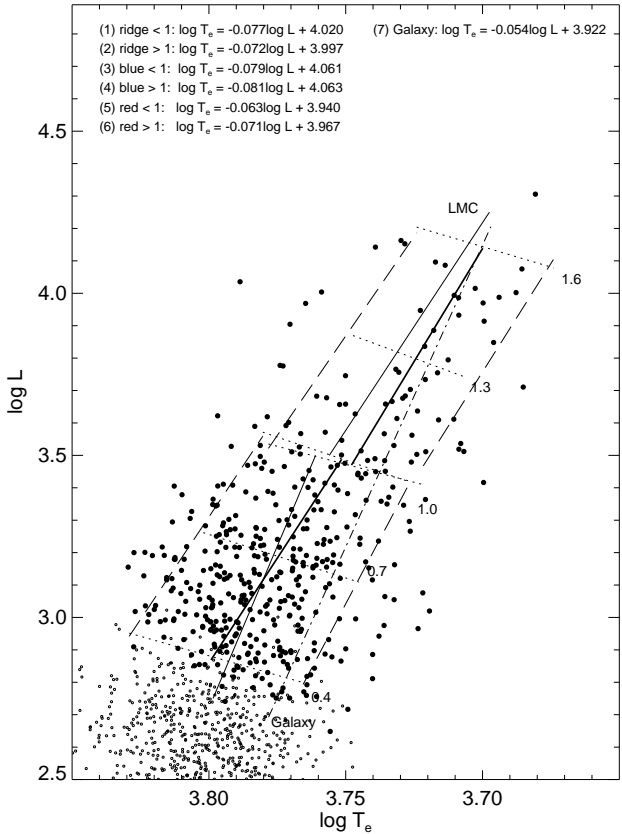}}
    \caption{The instability strip in $\log L_{V}$, $\log T_{e}$. The
   ridge-line relations for the Galaxy (dot-dashed line) and for the
   LMC and SMC with breaks at 10 days (solid lines). The individual
   SMC Cepheids are the dots. Cepheids with periods less than 
   $\log P=0.4$ are the small open circles. The equation of the SMC
   ridge lines are in the upper left part of the diagram.}
    \label{fig:08}
\end{figure}

\begin{figure}
   \centering
   \resizebox{\hsize}{!}{\includegraphics{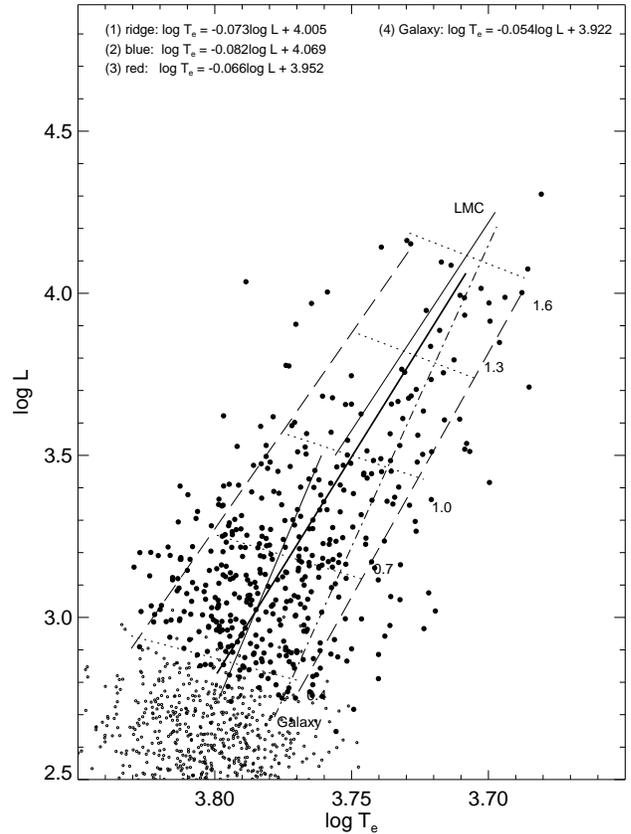}}
    \caption{Same as Fig.~\ref{fig:08} but showing the no-break
   solution for the SMC.}
    \label{fig:09}
\end{figure}

     The results for the SMC break case are shown in
Fig.~\ref{fig:08}. Those for the no-break solution are in
Fig.~\ref{fig:09}. Also shown are the ridge-line solutions for the
Galaxy (dot-dashed line) and the LMC (the heavy solid lines).       

% ********************************************************
% Figure 8: SMC - L Te with break
% ********************************************************
% ********************************************************
% ********************************************************
% Figure 9: SMC - L Te no break
% ********************************************************
% ********************************************************

     Both the SMC and the LMC ridge lines are at higher 
temperatures than the Galaxy line, and each have different slopes 
(except for LMC with $P<10$ days where the ridge line is nearly
parallel to that of the Galaxy). The consequences for these 
different slopes and zero points in the instability strips for
differences in the slopes and zero points of the respective P-L
relations are discussed in detail elsewhere (\citeauthor{ST:08}) and
are not repeated here. However, the consequences for the P-L relation
are shown graphically in the next section.

% ******************************************************************
% 7. SMC P-L Relation Compared with Those in the Galaxy, NGC 3351 
%    plus NGC 4321, the LMC, and NGC 3109 
% ******************************************************************
\section{SMC P-L relation compared with those in the Galaxy, NGC\,3351 
         plus NGC\,4321, the LMC, and NGC\,3109}
\label{sec:7}
\begin{figure}
   \centering
   \resizebox{\hsize}{!}{\includegraphics{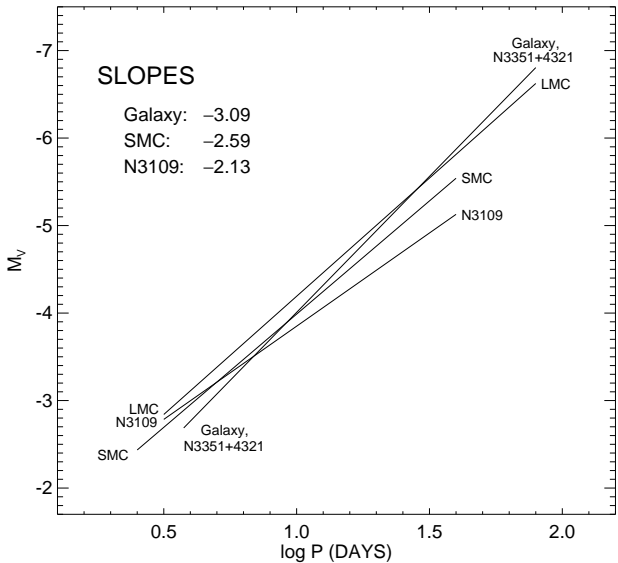}}
    \caption{Summary of the ridge lines of the P-L relations in 6
   galaxies including the data analyzed here for the SMC. The slope
   differences are evident. They become flatter as the metallicity of
   the host galaxy becomes smaller. The P-L relations shown for SMC
   and LMC are single-fit solutions with no allowance for a break at
   $P=10^{\rm d}$.}  
   \label{fig:10}
\end{figure}

To emphasize that the data for a number of galaxies now support the
proposition that the slopes of the Cepheid P-L relations differ from
galaxy-to-galaxy as function of metallicity (low metallicity Cepheids
have flatter P-L slopes), Fig.~\ref{fig:10} shows a summary of the P-L
relations of five galaxies (Galaxy, NGC\,3351 plus NGC\,4321, LMC, and
NGC\,3109) compared with the SMC here, illustrating the trend. 
More complete discussions are given elsewhere (\citeauthor{TSR:08}, 
Table~5 and Fig.~4; \citeauthor{ST:08}, Fig.~4).      

% ********************************************************
% Figure 10: SMC - PL comparison
% ********************************************************
% ********************************************************

     This diagram shows the importance of the SMC data because 
the difference between the Galaxy (or NGC\,3351 plus NGC\,4321) and 
the LMC is so small that a case made only using them is not 
convincing. However, the SMC P-L slope difference with LMC is 
large, and is supported more strongly when NGC\,3109 is added. 

     The point to again be made is that the instability strip 
differences in slope of the ridge lines in Fig.~\ref{fig:08} and
\ref{fig:09} must show as a difference in the slopes of the P-L
relations, as seen in Fig.~\ref{fig:10}. The case is supported because
the instability strip data in Figs.~\ref{fig:06}--\ref{fig:09} have
not been used to obtain the slopes of the P-L relations in
Fig.~\ref{fig:10}. These have been determined from the observational
data alone using only periods and apparent magnitudes of the Cepheids
in each galaxy.

% ******************************************************************
% 8. Conclusions
% ******************************************************************
\section{Conclusions}
\label{sec:8}
The LMC and SMC are the two galaxies for which the largest numbers of
classical Cepheids are known with excellent and consistent $BVI$
photometry \citep{Udalski:etal:99a,Udalski:etal:99b}. They yield
therefore particularly well to an intercomparison of their Cepheids,
which is favored in addition by only small reddenings. The P-C and P-L
relations of the Cepheids in the two galaxies,
although sharing the change of slope at $P=10\;$days,
are found to be significantly different in the overlapping period
range of $\log P>0.4$. 
At least one parameter which governs these differences
is the different metallicity of SMC ([O/H]=7.98) and LMC ([O/H]=8.34)
by a factor of 2.3. As confirmed also in other galaxies, metal-poor
Cepheids are bluer than metal-rich Cepheids not only because of their
weaker line blanketing effect, but also because of their higher
temperature. In addition metal-poor Cepheids tend to be brighter at
short periods and fainter at long periods than their Galactic
counterparts. These differences are more pronounced at shorter 
wavelengths. 

     The P-L relation of SMC predicts fainter luminosities than that
of LMC by about $0.2\mag$. For distance determinations the effect is
compensated or even overcompensated by the higher absorption
attributed to the {\em blue\/} SMC Cepheids, such that a Cepheid
sample with wide period coverage yields quite similar distances from
the SMC or LMC P-C and P-L relations. The SMC distance, however, is
{\em larger\/} at a period of 10 days by $0.1\mag$. Of course, for
still higher metallicities the difference can be larger for specific
periods (see \citeauthor{TSR:08b}, Table~3).

     Cepheid distances, consistently reduced as to metallicity and
period, are compiled in \citeauthor{TSR:08b}, Table~2. The resulting
``long'' distance scale is in excellent agreement with the independent
tip of the red-giant branch (TRGB) distance scale
(\citeauthor{TSR:08b}), demonstrating the necessity of metal-dependent
P-C and P-L relations if they are used for distance determinations.

     The existence of a universal P-L relation of classical
Cepheids is an only historically justified illusion
\citep[see also][]{Romaniello:etal:08}.

% ******************************************************************
%  Acknowledgments
% ******************************************************************
\begin{acknowledgements}
We thank Dres. Alfred Gautschy and Francoise Combes for helpful
informations and comments. A.\,S. thanks the Carnegie Institution for
post retirement support facilities. 
\end{acknowledgements}

% ******************************************************************
% Bibliography
% ******************************************************************

% ******************************************************************

% ******************************************************************
\end{document}